\documentclass{INTERSPEECH2023}

\interspeechcameraready

\title{QV\MakeLowercase{oice}: Arabic Speech Pronunciation Learning Application}

\name{Yassine El Kheir, Fouad Khnaisser, Shammur Absar Chowdhury, \\ Hamdy Mubarak, Shazia Afzal, Ahmed Ali}
\address{
  Qatar Computing Research Institute, HBKU, Doha, Qatar}
\email{qvoice@hbku.edu.qa}
\begin{document}

\maketitle
 
\begin{abstract}

This paper introduces a novel Arabic pronunciation learning application \textbf{QVoice}, powered with end-to-end mispronunciation detection and feedback generator module. The application is designed to support non-native Arabic speakers in enhancing their pronunciation skills, while also helping native speakers mitigate any potential influence from regional dialects on their Modern Standard Arabic (MSA) pronunciation.
QVoice employs various learning cues to aid learners in comprehending meaning, drawing connections with their existing knowledge of English language, 
and offers detailed feedback for pronunciation correction, along with contextual examples showcasing word usage.
The learning cues featured in QVoice encompass a wide range of meaningful information, such as visualizations of phrases/words and their translations, as well as phonetic transcriptions and transliterations. QVoice provides pronunciation feedback at the character level and assesses performance at the word level.

\end{abstract}

\section{Introduction}
Historically, acquiring a new language or improving pronunciation skills often demanded substantial 
effort and resources, typically in a classroom setting. Nevertheless, advancements in technology have paved the way for more personalized and self-paced learning experiences for language learners.
In recent years, the use of self-assessment tools has gained popularity as they provide learners with a means to track their progress and identify areas for improvement. Such tools are particularly useful in language learning, where children require extensive training to develop reading and pronunciation skills, on the other hand, non-native speakers often struggle with differences in their native language, furthermore, native speakers may face challenges with dialectal variations. 

This paper presents \textbf{QVoice}\footnote{https://apps.apple.com/gb/app/qvoice-apl/id6444646358} application which aims to provide learners with a comprehensive tool for training themselves to pronounce Arabic accurately and effectively; and demonstrates the potential of self-assessment tools in language learning. 

\section{QVoice Application}
\begin{figure*} [!ht]
\centering

\includegraphics[width=0.95\textwidth]{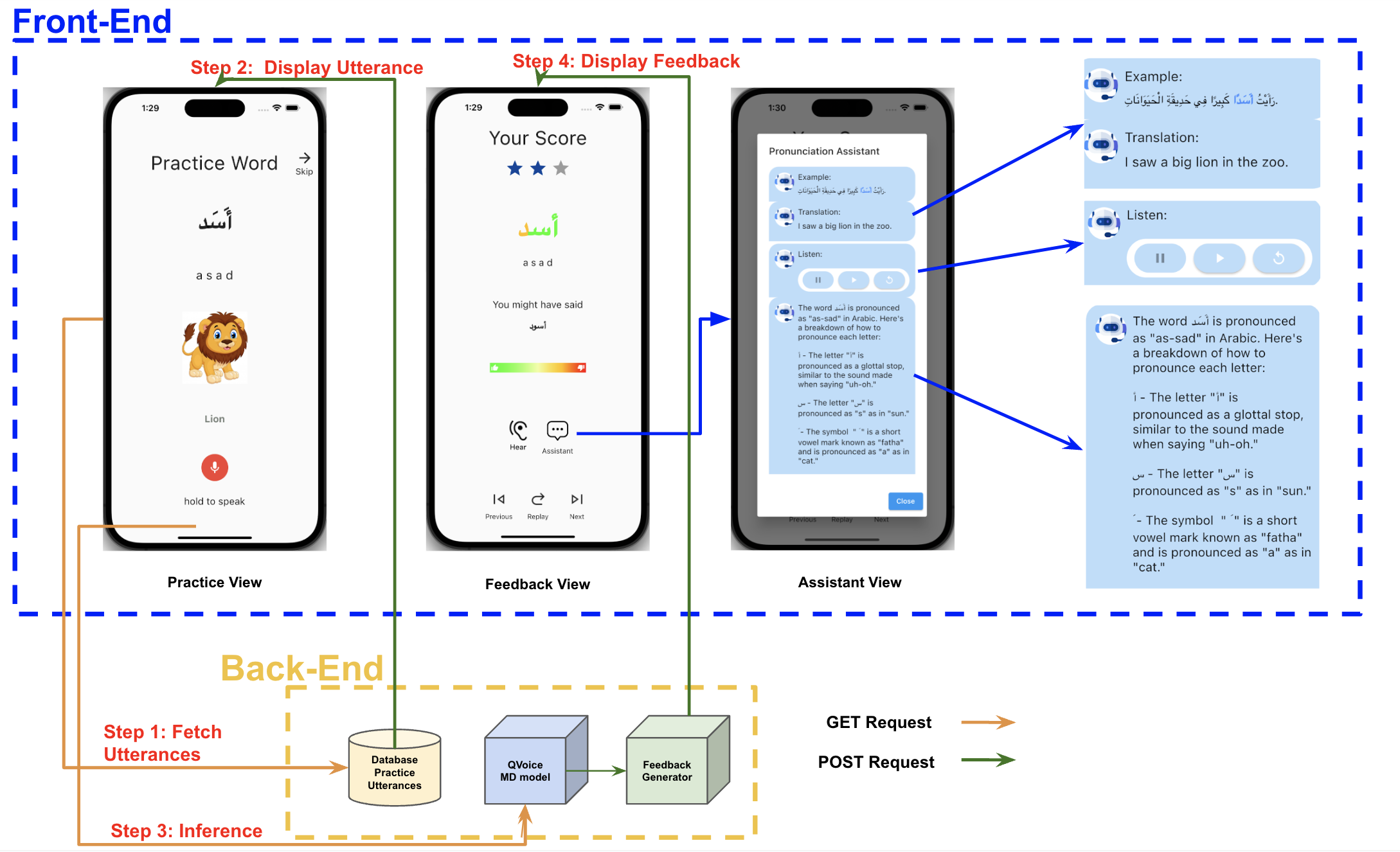}
\caption{\textbf{QVoice} Mobile App Font-End and Back-End}
\label{fig:APP}
\end{figure*}

The \textbf{QVoice} mobile application architecture comprises of a front-end and a back-end modules, as depicted in Figure \ref{fig:APP}. The front-end serves as the primary interface for learners, allowing to initialize their microphone, record their audio input and get feedback from the system.
We initialize the front-end by fetching the right word along with the 
additional learning cues: text, audio, and images, from the database. These words are then displayed to the user for practice. 
Once the user records their audio input, the data is sent to the back-end built using \textbf{Flask API}\footnote{https://flask-api.github.io/flask-api/} for processing and generating feedback. The generated feedback and supportive learning materials are then displayed to the learner in the front-end.



\subsection{End-to-End Mispronunciation Detection Model}
The Qvoice application uses an end-to-end mispronunciation detection model to predict pronunciation correctness. 
The system takes the input audio and estimates the score at the character level, alongside an overall utterance score to quantify the degree of mispronunciation. 
The model's scores enable the identification of insertions, deletions, substitutions, and mispronounced characters. Moreover, the end-to-end model incorporates an additional attention mechanism to detect mispronunciations due to dialectal influence. The proposed model has shown promising results in identifying various mispronunciation mistakes and can be useful for language learners and instructors.


\subsection{Database}
The current database contains more than $400$ words/phrases selected from multiple educational sources such as Aljazeera (\url{https://learning.aljazeera.net}), BTEC (Basic Travel Expression Corpus), etc. 
Each of these words/phrases is associated with a unique identifier, along with its corresponding transliteration, translation, reference audio exhibiting good pronunciation, an Arabic sentence example, the corresponding sentence translation, and additional information generated using an internal generative model that has been verified for each utterance. Additionally, each utterance was vowelized using \textbf{Farasa} \cite{darwish2017arabic}, and has an audio clip associated with it that has been generated using a text-to-speech (\textbf{TTS})\footnote{https://tts.qcri.org/reference/} engine \cite{abdelali2022natiq}. When a fetch request is received, a random word/phrase is selected 
and transmitted within less than one second.

\subsection{User Interface}
The user interface of QVoice includes two primary pages, designed using \textbf{Flutter} \footnote{https://flutter.dev}. The first page (Practice View) displays the targeted word and additional cues to the learners and allows them to record the pronunciation of the given word/phrase. Once recorded, the application transitions to the second page (Feedback View) which provides generated feedback to the learner. The details are given below: 

\paragraph*{Practice View:}  In this task view, the selected Arabic word is displayed to the learner, along with an image that depicts the meaning of the word, English transliteration and English translation. The recording can be initiated by pressing and holding the record button, which will begin pulsating to indicate that the recording has started. Upon completion, the button is released, and the feedback view displays the results fetched from our back-end end-to-end model in less than $1.5$ seconds.


\paragraph*{Feedback View:} The feedback page is designed to give fine-grained information about the pronunciation and puts it in context with L$1$ and L$2$ language. At the top of the page, an overall score at the utterance level is presented using a star-based rating system.
Following, the practiced example is shown, color-coded to signal the pronunciation accuracy at the character level. The colored bar portrays the significance of each color to its score with red means very poor to green representing excellent pronunciation.
The feedback view also presented the transliteration in English along with a predicted sequence from the model analysing what has been said by the user highlighting any characters or sounds that were omitted, added or mispronounced.
To aid the learner with audio feedback, the app provides a reference for pronunciation with different speed (normal and slowed reference). The app also supports an assistant view putting the practiced word in context with further information.

\paragraph*{Assistant View:} In the assistant view, the learner has access to a grammatically correct example containing the practiced word highlighted in blue. Along with the Arabic example, the assistant also provided its English translation and the Arabic-generated sample, using TTS engine. This audio feedback showcase how the practice word fits in a larger sentence/context. The learner can play, pause, and replay the audio as much as needed. Finally, a detailed graphophonic explanation is provided along with reference pronunciation with English examples to aid in further improving the user's pronunciation skills.


\section{Conclusion}
The proposed application is a fully functional Arabic pronunciation learning framework. The application facilitates the learners to practice Arabic pronunciation on their own time and pace. The fine-grained pronunciation feedback locates the position of the mispronunciation, and a combination of auditory, visual, and textual learning cues enables the learner to practice efficiently while exploiting prior L$1$ language knowledge. In future, we plan to extend the demo to incorporate articulatory features among other cues.
The practice words are selected from multiple scenarios. The presented system is modular and allows easy adaption to the desired content.
The back-end systems have high accuracy for detecting mispronunciation, the audio feedback/learning cues are generated using state–of–the–art TTS allowing natural responses.

\bibliographystyle{IEEEtran}
\bibliography{mybib}

\end{document}